\documentclass[aps,prl,showkeys,twocolumn,nofootinbib]{revtex4-1}
\usepackage{natbib}
\usepackage{amssymb}
\usepackage{color}
\newcommand{\ped}[1]{\ensuremath{_{\rm #1}}}
\newcommand{\apex}[1]{\ensuremath{^{\rm #1}}}
\usepackage{float}
\usepackage[caption = false]{subfig}
\usepackage{graphicx}
\definecolor{blue}{rgb}{0,0,0}

\begin{document}

\title{Anomalous screening of an electrostatic field at the surface of niobium nitride}

\author {Erik Piatti}
\author {Davide Romanin}
\author {Renato S. Gonnelli}
\author {Dario Daghero}
\email{dario.daghero@polito.it}
\affiliation {Department of Applied Science and Technology, Politecnico di Torino, corso Duca degli Abruzzi 24, 10129 TO Torino, Italy}

\begin{abstract}
The interaction between an electric field and the electric charges in a material is described by electrostatic screening, which in metallic systems is commonly thought to be confined within a distance of the order of the Thomas-Fermi length. The validity of this picture, which holds for surface charges up to $\sim10^{13}$ cm\apex{-2}, has been recently questioned by several experimental results when dealing with larger surface charges, such as those routinely achieved via the ionic gating technique. Whether these results can be accounted for in a purely electrostatic picture is still debated. In this work, we tackle this issue by calculating the spatial dependence of the charge carrier density in thin slabs of niobium nitride via an \emph{ab initio} density functional theory approach in the field-effect transistor configuration. We find that perturbations induced by surface charges $\lesssim 10^{14}$ cm\apex{-2} are mainly screened within the first layer, while those induced by larger surface charges $\sim10^{15}$ cm\apex{-2} can penetrate over multiple atomic layers, in reasonable agreement with the available experimental data. Furthermore, we show that a significant contribution to the screening of large fields is associated not only to the accumulation layer of the induced charge carriers at the surface, but also to the polarization of the pre-existing charge density of the undoped system.
\end{abstract}
%

\maketitle

\section{Introduction}
The electric field effect (i.e., the modulation of the conduction properties of a material by means of an electrostatic field) is well known and widely used in today's electronic components like field-effect transistors (FET). With respect to conventional solid-state FETs, the ionic gating technique (based on the use of an electrolyte instead of the solid dielectric) allows an enhancement up to two orders of magnitude in the field intensity, thanks to the formation of an electric double layer (EDL) at the interface between the electrolyte and the material under study, which acts as a capacitor with a nanometric inter-plate spacing and a huge capacitance \cite{UenoReview2014}. In the past few years, this technique has led to the discovery and control of new phases (including the superconducting one) in various materials. In their native state, most of these featured a low to moderate carrier density ($\lesssim 10^{14}$ cm\apex{-2}), and ranged from insulating oxides \cite{UenoNatMater2008,UenoNatNano2011,JeongScience2013,LengQMater2017}, to various layered materials \cite{YeScience2012, JoNanoLett2015, YuNatNano2015, ShiSciRep2015, GonnelliSciRep2015, SaitoScience2015, SaitoACSNano2015, PiattiAPL2017, PiattiarXiv2018}, to cuprate superconductors \cite{BollingerNature2011,LengPRL2011,LengPRL2012,ZhangACSNano2017}. A more limited attention has been paid instead to materials with a large intrinsic carrier density, where the effect of the field was generally thought to be undetectable because of the strong electrostatic screening. We have shown however \cite{DagheroPRL2012,TortelloApsusc2013} that, even in thin films of noble metals, a measurable modulation of the resistivity (up to 10\% at low temperature) can be induced by ionic gating. This was followed by reports on the successful modulation of the superconducting properties of standard BCS superconductors \cite{ChoiAPL2014,PiattiJSNM2016,PiattiPRB2017}, metallic transition-metal dichalcogenides \cite{LiNature2016,XiPRL2016,WuAPL2018} and iron-based compounds \cite{ShiogaiNatPhys2016,LeiPRL2016,ZhuPRB2017}.

In many of these experiments, the thickness of the charge accumulation layer was found to increase with the intensity of the field (eventually becoming much larger than the Thomas-Fermi value $\lambda_{TF}$) in stark contrast with the expectation that the screening length should decrease on increasing the carrier density. This anomaly was observed in the inversion layer of insulating SrTiO\ped{3} \cite{UenoPRB2014,ValentinisPRB2017}, in ultra-thin films of YBa\ped{2}Cu\ped{3}O\ped{7-x} \cite{FeteAPL2016}, and in thin films of the standard electron-phonon superconductor NbN \cite{PiattiJSNM2016,PiattiPRB2017}. In all cases, this anomalous behaviour was observed in superconducting systems and for large values of the surface density of induced charge, i.e. $\Delta n_{2D}\gg 10^{13}$ cm\apex{-2}. Different explanations of this phenomenon have been proposed, sometimes involving peculiarities of the material under study (i.e. incipient ferroelectricity in SrTiO\ped{3} \cite{UenoPRB2014}, or redistribution of oxygen atoms in YBCO \cite{FeteAPL2016}). The ubiquity of the effect, also observed in a simple system like NbN \cite{PiattiJSNM2016,PiattiPRB2017} rather suggests a general phenomenon.
Clearly, the increase in the screening length indicates the breakdown of the linear Thomas-Fermi approximation \cite{PiattiJSNM2016} in the presence of ultra-high electric fields. In Ref. \citenum{PiattiPRB2017}, we used the theory for electrostatic screening beyond the linear regime \cite{ChalzavielBook}, in the non-linearized Thomas-Fermi approach, and we showed that this approach is able to reproduce the experimental data only up to a limited $\Delta n_{2D}\sim 5\cdot10^{14}\,\mathrm{cm}\apex{-2}$.

Therefore, the issue remains open: is there a basic mechanism (not necessarily involving material-specific features) that can explain the observed increase in the thickness of the accumulation layer in very high fields, where the linear approximation breaks down? To understand whether this is the case, a more comprehensive modellization of the material response to an ultra-high electric field is necessary. {\color{blue}Such a modellization must clearly go beyond the Thomas-Fermi approximation, whose applicability to a quantitative description of electrostatic screening in metallic systems is in any case questionable. As a matter of fact, this approximation holds for slowly varying potentials (i.e. potentials varying on a length scale larger than  $\lambda_F$) and is intrinsically based on the hypothesis of a uniform positive background \cite{ChalzavielBook}. Considering that in NbN  $\lambda_F \simeq 4$ {\AA} \cite{ChockalingamPRB2008} and the interplanar distance is $2.21$ {\AA}, the calculated  $\lambda_{TF} \simeq 1 $ {\AA} is clearly in contradiction with both the aforementioned basic assumptions. 

In this paper we calculate the response of a metallic material to an external (static) electric field, without any specific assumption. To do so, we use a first-principles approach and a model systems that mimics the configuration used for field-effect experiments.} To allow a direct comparison with experimental results, we chose NbN as the material under study.

\section{Computational details}
In order to study the screening in NbN, we used the \emph{ab initio} approach described in Refs.\citenum{BrummePRB2014, BrummePRB2015} designed to mimic the configuration of a real EDL field-effect experiment, in which the electric field is created by a planar distribution of charges (the ions in the liquid gate or polymer-electrolyte solution) in close proximity of the surface of a thin film, with no solid dielectric in between.

\begin{figure}[h]
\begin{center}
\includegraphics[keepaspectratio, width=\columnwidth]{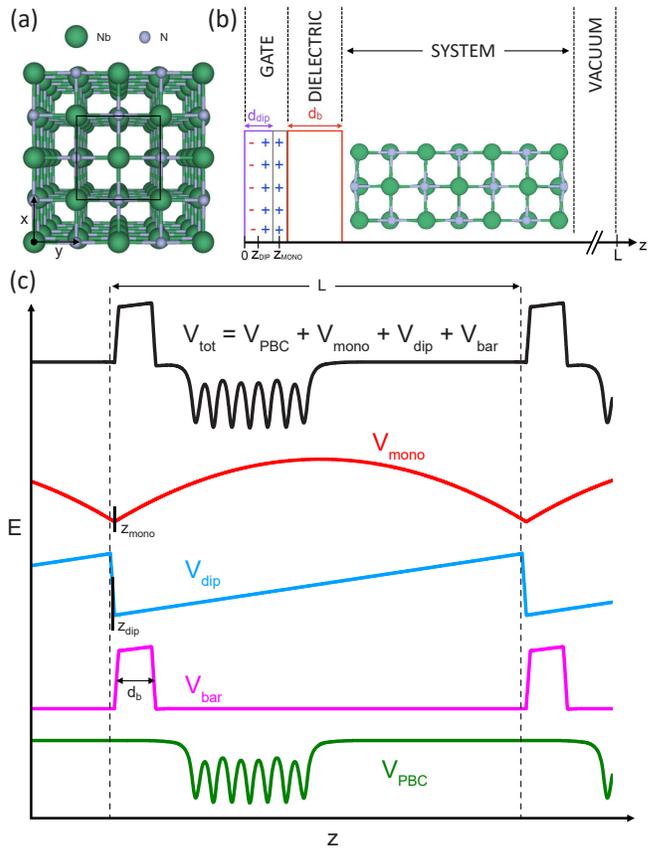}
\end{center}
\caption {
(a) Ball-and-stick model of the NbN lattice in perspective top view. Large green balls are niobium. Small grey balls are nitrogen. The unit cell is indicated by a solid black line. (b) Structure of the NbN $7$-layer system in the FET configuration (side view). {\color{blue}(c) Schematic picture of the planar averaged Kohn-Sham potential (V\ped{tot}, black line) for periodically repeated, charged slabs, in the absence of the exchange-correlation potential. Its components (not in scale) are highlighted with different colors: monopole potential (V\ped{mono}, red line), dipole potential (V\ped{dip}, blue line), potential barrier (V\ped{bar}, magenta line), and bare ionic potential as obtained from the periodic boundary conditions (V\ped{PBC}, green line).}} \label{figure:layout}
\end{figure}

To model the NbN thin film, we considered a slab structure made of $7$ atomic layers. We initially built the structure by using the lattice constant $a=4.38$ {\AA} determined from neutron scattering experiments \cite{Chen2005}, that became $a=4.42$ {\AA} after bulk relaxation.

Let us call $z$ the axis perpendicular to the layers. The ions accumulated on the surface of the film are modeled with a sheet of uniformly distributed charges, in the following called monopole, placed at $z_{mono}$. The surface density of charges $\sigma$ in the monopole corresponds to the experimentally-accessible surface density of induced charges, i.e. $\sigma = e\Delta n_{2D}$. An equal amount of opposite charge is given to the lattice to mimic the fact that, in an EDL experiment, the film is not floating but connected to the measurement circuit. The dielectric is completely removed and substituted with a uniform potential barrier of width $d_b$ and height $6.0\,\mathrm{Ry}$. This potential barrier is necessary because, when relaxing the structure under an applied field, one wants the ions not to move too close to the charged plate.
To be able to work under periodic boundary conditions (PBC), we included $\sim 20\,\mathrm{\AA}$ of vacuum between two successive repeated images of the system to avoid any unphysical interaction between them. Since in this vacuum region the field needs to be zero, we placed a dipole (generated by two planes of opposite charge at a distance $d_{dip}$) in the vacuum region next to the monopole, at $z=z_{dip}$. The supercell including the dipole, the monopole, the lattice and the vacuum has length $L$ and we fixed $z_{dip}=0.005\,L$, $d_{dip} = 2 z_{dip}=0.01 \, L$, $z_{mono}=0.011\,L$, and $d_b=0.1 \,L$.A sketch of the configuration is given in Fig.\ref{figure:layout}b and a picture of the $z$-dependence of the various potentials is reported in {\color{blue}Fig.\ref{figure:layout}c. Further details about the model are discussed at length in Ref.\citenum{BrummePRB2014}.}

Once set the model, we performed \textit{ab initio} computations of the charge density in the lattice by using density functional theory (DFT) within the Quantum Espresso package \cite{GiannozziQEspresso}, which is based on the expansion of valence-electron wave functions and charge density in terms of plane waves. In order to model the exchange-correlation functional we used a revised Perdew-Burke-Ernzerhof generalized gradient approximation (PBEsol \cite{PerdewPRL2009}), which is particularly suited for densely packed solids and their surfaces.  The core-electrons contribution of Nb was approximated with a projector-augmented wave pseudopotential (PAW \cite{BlochlPRB1994}) while that of N with an ultrasoft pseudopotential \cite{VanderbiltPRB1990}, both obtained from the SSSP library \cite{SSSP}. We set the cut-off for the expansion of valence-electron wave functions to $65\,\mathrm{Ry}$ and that for the density to $750\,\mathrm{Ry}$. We performed the Brillouin zone integration with a Monkhorst-Pack grid of $32\times32\times1$ k-points both for the neutral and charged systems, with a Methfessel-Paxton first-order spreading \cite{MethfesselPRB1989} of $0.025\,\mathrm{Ry}$. We set the convergence criteria for the self-consistent solution of the Kohn-Sham equations to $10^{-9}\,\mathrm{Ry}$ for the total energy and to $10^{-3}\,\mathrm{Ry}/a_0$ (where $a_0$ is the Bohr radius) for the maximum force acting on atoms during the relaxation of the slab. Charge maps and structural drawings were obtained using the VESTA software \cite{Vesta}.

The simulations were performed as follows: i) first of all, the bulk NbN structure was relaxed without any external field by minimizing the total energy. This gave a lattice constant $a=4.42$ {\AA}; ii) Subsequently, the supercell was built in order to model the surface of the material, and the structure was further relaxed in zero external field; iii) Then, for every value of the surface charge density (i.e. of the electric field), the slab was relaxed in order to determine the structure corresponding to the minimum total energy in the presence of this anisotropic perturbation, and the equilibrium position of the surface with respect to the potential barrier. The equilibrium distance between the first layer and the potential barrier turned out to be $\simeq 5$ {\AA} for all the field intensities analyzed here. Note that all these relaxations were performed in the harmonic approximation of the ionic potential: in principle, this may be an issue for the computation of the phonon bandstructure of the slab, since the stoichiometric rocksalt phase of bulk NbN is unstable, leading to the presence of imaginary phonons \cite{BlackburnPRB2011}. While this instability may be lifted by taking into account anharmonic effects \cite{BlackburnPRB2011}, we neglected these corrections in this work, as the current version of the Quantum Espresso package (v.6.2.1) is not able to compute the vibrational properties of a material in the FET configuration; iv) Once the relaxed lattice structure was determined for each field intensity, the charge density was obtained by the self-consistent solution of the Kohn-Sham equations for both the neutral and charge-doped slabs, using the relaxed structure of the latter.

\section{Results}

\begin{figure}[t]
\begin{center}
\includegraphics[keepaspectratio, width=0.95\columnwidth]{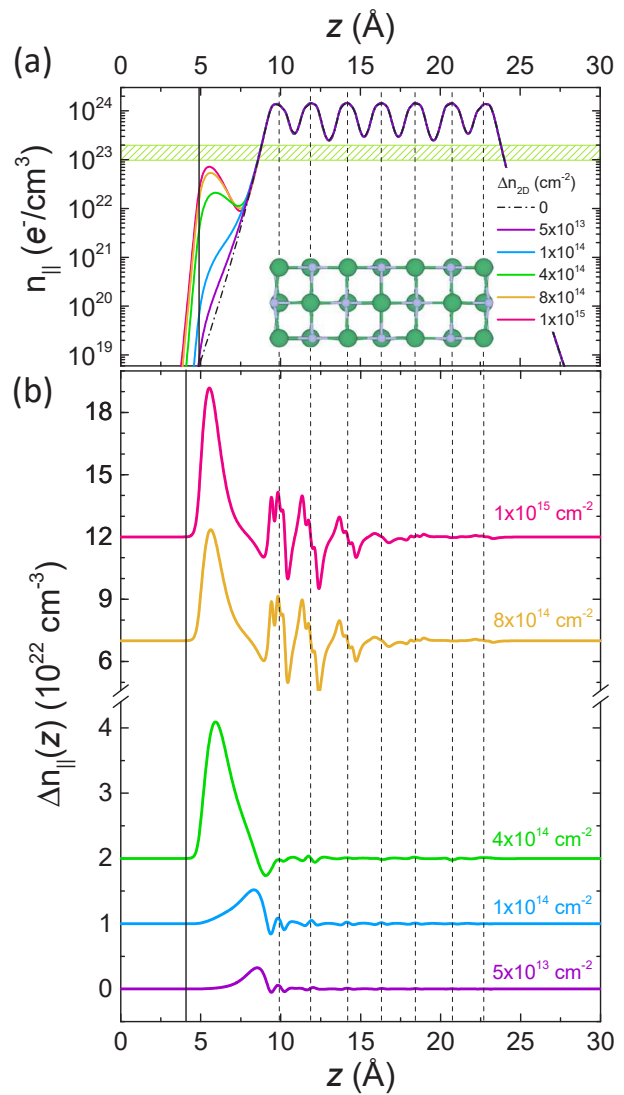}
\end{center}
\caption {
(a) $z$-dependence of the planar average of the total electronic charge in the NbN $7$-layer system, $n_{||}$, for different values of induced charge density $\Delta n_{2D}$. The green shaded band represents the typical free carrier density values for NbN thin films \cite{ChockalingamPRB2008}. The vertical solid line indicates the position of the potential barrier; the position of the NbN slab with respect to the barrier comes from the relaxation of the supercell in the presence of the field. (b) Planar averaged difference in the full charge density between the doped and undoped slab, $\Delta n_{||}$, for different values of induced charge density $\Delta n_{2D}$. {\color{blue}Each curve is vertically offset for clarity (vertical steps are $1\cdot10^{22}\,\mathrm{cm^{-3}}$ for the three bottom curves and $5\cdot10^{22}\,\mathrm{cm^{-3}}$ for the two top curves).} Vertical dashed lines mark the positions of the atomic planes.} \label{figure:integrated_profiles}
\end{figure}

Let us first consider how the application of the electric field alters the \emph{total} charge density in the NbN slab. In Fig.\ref{figure:integrated_profiles}a, we plot the $z$-axis dependence of the total electron density, averaged along the ($x$,$y$) plane ($n_{||}$). Note that this includes both the contribution of conduction and valence electrons. In the absence of an applied electric field (black dash-dotted line), the electron density is symmetric with respect to the central (fourth) layer of the relaxed slab; electrons are well confined close to the atomic positions, with $\sim 99$\% of the charge density within $\sim 3\,\mathrm{\AA}$ of the outermost layers. The remaining charge density extends in the evanescent tails of the wavefunctions, that eventually reach the gate potential barrier located $\sim 5\,\mathrm{\AA}$ away from the first atomic layer. The application of an electric field breaks the symmetry of the system and leads to the formation of an electron accumulation layer close to the surface (solid lines), which gives the predominant contribution to the doping charge $\Delta n_{2D}$. This accumulation layer does not form in correspondence of the first atomic layer, but instead develops from the evanescent tails of the pristine electron density. Note that, for the largest $\Delta n_{2D} = 1\cdot10^{15}$ cm\apex{-2}, the maximum value of $n_{||}$ in the accumulation layer is only slightly lower than the average free carrier density for NbN, as determined in thin films by Hall-effect measurements within the single-band model \cite{ChockalingamPRB2008} (green shaded band).

Fig.\ref{figure:integrated_profiles}b shows the difference $\Delta n_{||}$ between the planar average of the total electron density in the doped NbN slab, $n_{||}(\Delta n_{2D}\neq0)$, and in the undoped slab, $n_{||}(\Delta n_{2D}=0)$. The latter is calculated with the relaxed lattice structure of the former. This difference represents the planar average of the screening charge, including the contributions from both the free electrons in the conduction band and the dielectric response of the rearranged valence electrons \cite{BrummePRB2014}. Negative values of $\Delta n_{||}$ obviously indicate electron depletion, and positive values electron accumulation. We observe that the electric field does not only produce the emergence of the accumulation layer at the surface, but also induces a polarization of the charge density in the first few atomic layers, represented by the oscillations in $\Delta n_{||}$ within the slab.
For moderate electric fields ($\Delta n_{2D}\lesssim1\cdot10^{14}$ cm\apex{-2}), these oscillations are limited in intensity and mostly confined to the first atomic layer. For large electric fields ($\Delta n_{2D}\gtrsim8\cdot10^{14}$ cm\apex{-2}), the oscillations extend deeper in the slab and their amplitude is significant up to the third atomic layer; In correspondence of the first two layers (first unit cell) the local screening-charge density $\Delta n_{||}$ can be as large as $\sim 30$\% of the maximum. Additionally, on increasing $\Delta n_{2D}$ the position of the maximum density in the accumulation layer shifts away from the first atomic layer and localizes in close proximity to the potential barrier{\color{blue}, while the width of the accumulation layer slightly decreases. This latter observation suggests that the effective ``screening length" \emph{within the accumulation layer} decreases for higher electric fields, as is normally expected. However, the screening it provides becomes incomplete, resulting in the polarization of the charge density in the underlying atomic layers.}

\begin{figure}[t]
\begin{center}
\includegraphics[keepaspectratio, width=\columnwidth]{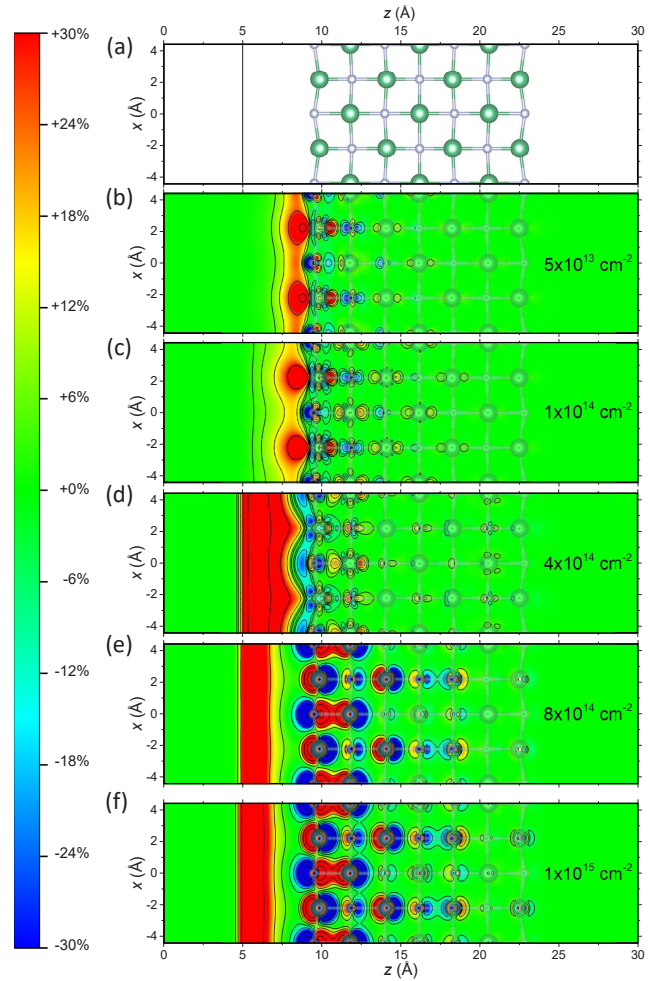}
\end{center}
\caption{
(a) Ball-and-stick model of the ($x$,$z$) plane of the NbN lattice. The vertical solid line indicates the position of the potential barrier. (b-e) Difference between the total charge density in doped and undoped NbN slab, sliced along the ($x$,$z$) plane. Color maps are in linear scale between $\pm30$\% of the maximum density difference in the accumulation layer. The lattice structure is shown in the background of each panel.} \label{figure:color_maps}
\end{figure}

Further insight into the spatial distribution of the screening charge can be obtained by dropping the planar averaging, and considering two-dimensional maps along specific crystallographic planes. Hence, we directly calculated the difference $\Delta n_{3D}$ between the total electron density in the doped and undoped NbN slab (again, the latter was obtained with the relaxed lattice structure of the former). Note that the integral of $\Delta n_{3D}$ along the $z$ axis must necessarily be equal to $\Delta n_{2D}$. In Fig.\ref{figure:color_maps} we show planar maps of $\Delta n_{3D}$ along the ($x$,$z$) plane, normalized to its maximum value in the accumulation layer. For moderate electric fields ($\Delta n_{2D}\lesssim1\cdot10^{14}$ cm\apex{-2}, panels b and c), the screening charge mainly accumulates in the $d_{z^2}$ orbitals of the outermost Nb atoms. Conversely, the $p_{z}$ orbitals of the outermost N atoms appear to be mostly interested by the formation of a charge dipole. A second row of dipoles forms between the Nb atoms in the first layer and the N atoms in the second layer. For large electric fields ($\Delta n_{2D}\gtrsim8\cdot10^{14}$ cm\apex{-2}, panels e and f), a well-defined accumulation layer forms, separated from the pristine Nb orbitals, and delocalized across the ($x$,$y$) plane. Furthermore, the formation of strong charge dipoles involves both Nb and N atoms at least into the third layer from the surface. Note that, in each ($x$,$y$) plane, the dipoles localized on the Nb atoms are in counterphase with those localized on the N atoms, leading to these contributions almost canceling each other out in the planar-averaged profiles. {\color{blue}This also suggests that these charge density oscillations cannot be simply interpreted as Friedel oscillations, which do not feature any dependency on the \emph{(x,y)} coordinates \cite{ChalzavielBook}.}

\section{Discussion}

Our result clearly show how, even in a metallic system such as NbN, the effects of intense electric fields extend well beyond the first atomic layer, and are not limited to pure surface accumulation of the extra induced carriers. Moreover, the charge accumulation layer extends even outside the material. Finally, the depth in the bulk over which the field is screened increases with increasing the field intensity. This last result is, at least qualitatively, in agreement with experimental findings in EDL-based field-effect measurements.
A quantitative comparison  with the experimental results is however not straightforward for two reasons. The first is that, as explained below, the outcome of experiments is an effective thickness of the accumulation layer that is hard to define starting from the results of DFT calculations. The second is that, in general, transport experiments (even in the superconducting state) probe only the conduction charge, i.e. the charge that falls within an energy window of width $k_B T$ around the chemical potential. Evaluating this charge, and especially determining where exactly it is located in direct space, is a difficult task in DFT, because the bandstructure of the surface is much more complicated than that of the bulk. The problem is particularly complex in the case of intrinsically metallic systems that possess mobile charges even in the absence of electric fields. This means that the $\Delta n_{||}$ profiles shown in Fig. \ref{figure:integrated_profiles} contain contributions from both the \emph{induced} charge and the \emph{intrinsic} charge.

\begin{figure}[t]
\begin{center}
\includegraphics[keepaspectratio, width=\columnwidth]{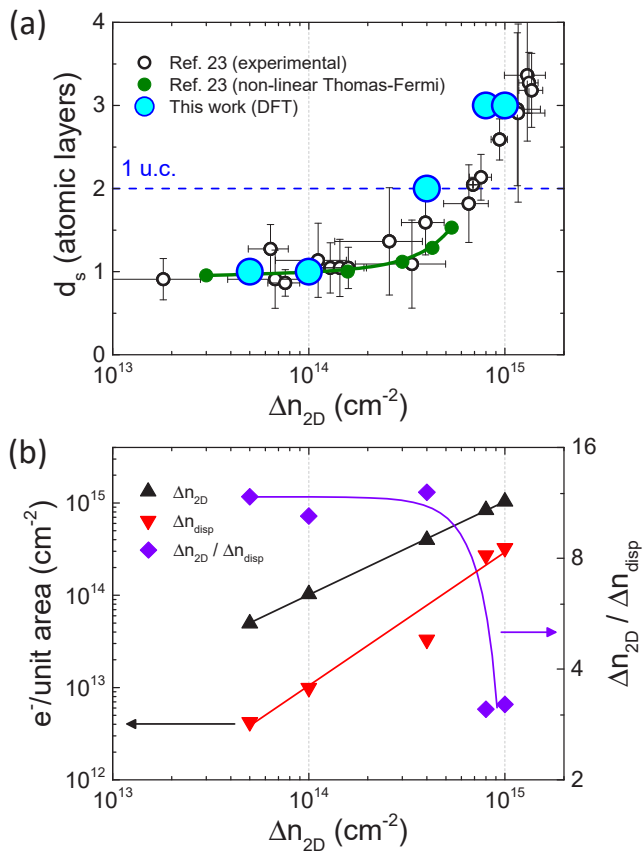}
\end{center}
\caption{
(a) Number of atomic layers involved in the screening of the electric field, as a function of the induced charge density. Filled blue dots are the results of the DFT calculations. Hollow black dots are taken from the effective screening length as determined experimentally \cite{PiattiPRB2017}. {\color{blue}Filled green dots are the screening length determined with a non-linearized Thomas-Fermi approach \cite{PiattiPRB2017}, and are shown for comparison.} (b) Contributions to the screening charge due to the induced (black {\color{blue}up triangles}) and displaced (red {\color{blue}down triangles}) charge densities, and their ratio (violet {\color{blue}diamonds}), as a function of the induced charge density. {\color{blue}Solid lines are guides to the eye.}}
\label{figure:screening}
\end{figure}

This said, we can propose a qualitative, preliminary comparison between the results of this study and the experimental observations, leaving a more detailed and quantitative treatment for future works.
The first issue is the determination of an effective screening length, $d_s$, from either the planar average of the charge densities in Fig.\ref{figure:integrated_profiles}b, or from the color maps in Fig.\ref{figure:color_maps}(b-d). The definition of this screening length should ideally be consistent with those from simpler models (such as the Thomas-Fermi approximation) and the quantities determined from the experiments (such as the effective thickness of the surface layer). This requirement is, however, rather problematic. The Thomas-Fermi length, $\lambda_{TF}$, assumes that the electric field decays in the material with a simple exponential dependence, while the Friedel oscillations follow a periodicity with the Fermi wavevector \cite{ChalzavielBook}. On the other hand, the surface layer thicknesses obtained from experiments are typically calculated assuming an even simpler step-like behavior for the density of (superfluid) charge carriers \cite{UenoPRB2014,ValentinisPRB2017,PiattiPRB2017}. Furthermore, all of these neglect a true atomistic description of the lattice structure and electron wavefunctions, which is instead included in the DFT results. Therefore, we decided to employ a simple heuristic criterion for the comparison: for each doping charge, we consider the $\Delta n_{3D}$ color maps, and consider an atomic layer to be perturbed if its electron density is enhanced by at least $30$\% of the maximum density in the accumulation layer. This criterion is somehow justified by the similarity with the definition of the characteristic decay length of an exponential behaviour. We then compare the number of perturbed layers with the experimental value of $d_s$ (extracted from transport measurements in NbN thin films \cite{PiattiPRB2017}) in units of the spacing between two layers in the bulk. As we show in Fig.\ref{figure:screening}a, we obtain a reasonable agreement between the calculations and the experimental values. While the agreement is clearly sound from a qualitative point of view, care has to be taken with assuming this to be a true \textit{quantitative} estimation: first, this heuristic comparison neglects the ``stretching'' of the accumulation layer towards the potential barrier for large fields; second, different choices of the perturbation threshold may lead to different estimations of the number of atomic layers involved. On the other hand, we stress that the experimental values are determined employing a model assuming a simple step-like behavior for the induced charge density \cite{PiattiPRB2017,UmmarinoPRB2017}, which is an even rougher approximation of the real system.

The second result that requires a more thorough quantification concerns the simultaneous induction of the accumulation layer and strong charge dipoles in the first few atomic layers, and how the relative intensity of these two contributions evolves with increasing electric fields. That is, the doped slab can be characterized not only by the additional countercharge $\Delta n_{2D}$, induced by the gate, but also by the amount of pristine charge that is displaced and participates in the polarization of the structure, $\Delta n_{disp}$. These can be obtained easily thanks to the charge conservation law. As already pointed out, the surface density of induced charge $\Delta n_{2D}$ (which is equal to the charge density in the monopole, $\sigma$, divided by the elementary charge $e$)  is simply the total integral along the $z$ direction of $\Delta n_{||}(z)$, the difference between the planar averages of the electron densities in the doped and undoped slab:
\begin{equation}
\int_{0}^{\infty}\Delta n_{||}(z)dz = \Delta n_{2D}
\end{equation}
since the planar average of the \emph{pristine} charge -- that is displaced in the formation of the dipoles, and that accounts for the oscillating behavior of $\Delta n_{||}(z)$ in the slab -- averages to zero when integrated. Hence, when the same integration is performed on the absolute value of $\Delta n_{||}(z)$, the displaced charge gives a contribution equal to twice $\Delta n_{disp}$, i.e. :
\begin{equation}
\int_{0}^{\infty}|\Delta n_{||}(z)|dz = \Delta n_{2D} + 2\Delta n_{disp}
\end{equation}
Thus, the ratio between the surface densities of \emph{induced} charge and \emph{displaced} intrinsic charge can be easily computed for any electric field intensity as:
\begin{equation}
\frac{\Delta n_{2D}}{\Delta n_{disp}} = \frac{2\int_{0}^{\infty}\Delta n_{||}(z)dz}{\int_{0}^{\infty}[|\Delta n_{||}(z)|-\Delta n_{||}(z)]dz}
\end{equation}

The values of $\Delta n_{2D}$ and $\Delta n_{disp}$ and their ratio $\Delta n_{2D}/ \Delta n_{disp}$ are plotted in Fig.\ref{figure:screening}b as a function of $\Delta n_{2D}$ (which is directly proportional to the electric field intensity at the surface). For small electric fields, $\Delta n_{2D}\gg\Delta n_{disp}$, and most of the perturbation to the system can simply be described by the formation of the surface accumulation layer. For large electric fields ($\Delta n_{2D}\gtrsim8\cdot10^{14}\,$cm\apex{-2}) $\Delta n_{disp}$ becomes comparable with $\Delta n_{2D}$, and a description of the perturbed system in terms of the presence of the accumulation layer alone is clearly no longer accurate. Both of these anomalies (the expansion of the perturbation across multiple atomic layers beyond the first unit cell, and the increasing relevance of the dipole charge) are likely to become even more pronounced for larger electric fields.

\section{Conclusions}
In summary, we have performed \textit{ab-initio} DFT calculations on a [100]-oriented NbN slab in the field-effect transistor configuration. We have determined how the spatial dependence of the electron density is changed by the combined effects of the applied electric field and induction of extra charge carriers, typical of field-effect experiments. The electric field is screened by the combination of the accumulation layer (due to the induced carriers) and the polarization of the pre-existent electron density. We observed that the screening charge is localized within the first atomic layer at low and moderate fields, but extends well beyond the first unit cells at high fields. This reproduces well the anomalous expansion of the perturbed surface layer reported in recent experimental literature. Additionally, we found that the formation of dipoles deep in the slab is negligible at low and moderate applied fields, but becomes relevant at high fields, involving a charge density per unit area comparable to that involved in the accumulation layer. Overall,  our result clearly establish how simplistic screening models, such as the often-used Thomas-Fermi approximation, do not offer a good approximation of a realistic treatment of electrostatic screening in the presence of large electric fields, even in normal metallic systems.

\section*{Acknowledgments}
We thank M. Calandra for fruitful scientific discussions. Computational resources were provided by hpc@polito (http://www.hpc.polito.it).


\end{document}